\documentclass[a4paper,11pt]{article}
\pdfoutput=1 

\usepackage{jheppub} 
\usepackage{graphicx}  
\usepackage{dcolumn}   
\usepackage{bm}        
\usepackage{amssymb}   
\usepackage{amsmath}
\usepackage{slashed}
\usepackage{epsfig}
\usepackage{comment}
\usepackage{epstopdf}
\usepackage[usenames,dvipsnames]{xcolor}
\usepackage{comment}
\usepackage[T1]{fontenc} 
\usepackage{color}
\begin{document}

\title{{\bf Covariant non-local action for massless QED and the curvature expansion}}

\author{John F. Donoghue}
\author{and Basem Kamal El-Menoufi}

\affiliation{Department of Physics,
University of Massachusetts\\
Amherst, MA  01003, USA}

\emailAdd{donoghue@physics.umass.edu}
\emailAdd{bmahmoud@physics.umass.edu}

\abstract{
We explore the properties of non-local effective actions which include gravitational couplings. Non-local functions originally defined in flat space can not be easily generalized to curved space. The problem is made worse by the calculational impossibility of providing closed form
expressions in a general metric. The technique of covariant perturbation theory (CPT) has been pioneered by Vilkovisky, Barvinsky and collaborators whereby the effective action is displayed as an expansion in the generalized curvatures similar to the Schwinger-De Witt local expansion. We present an alternative procedure to construct the non-local action which we call {\em non-linear completion}. Our approach is in one-to-one correspondence with the more familiar diagrammatic expansion of the effective action. This technique moreover enables us to decide on the appropriate non-local action that generates the QED trace anomaly in 4$D$. In particular we discuss carefully the curved space generalization of $\ln \Box$, and show that the anomaly requires both the anomalous logarithm as well as $1/\Box$ term where the latter is related to the Riegert anomaly action.}
\maketitle
\flushbottom
\section{Introduction}

While the fundamental Lagrangians describing our known physical theories are all local, quantum loops of massless or nearly massless particles yield non-local effects. It is often useful to arrange those loop effects into a non-local effective action which enables a systematic investigation of the quantum effects on the classical background fields. For theories where the symmetries relate the couplings of different types of particles, such as chiral theories or general relativity, the evaluation of a single loop using the background field method allows the loop corrections to a large number of processes to be calculated at once. For example in chiral perturbation theory, the renormalized non-local effective action \cite{Gasser} is useful for many different reactions.

In general relativity, Barvinsky, Vilkovisky and collaborators (hereafter referred to collectively as BV) have developed techniques for calculating and displaying the non-local gravitational effective actions that arise due to graviton loops or those of other massless fields \cite{Barvinsky1,Barvinsky2,Barvinsky3,Avramidi1,Avramidi2,Avramidi3,gospel,Gusev,Vilkovisky,Codello}. The results are presented using an expansion in the curvature. In effective field theory we are used to an expansion in the curvature for local Lagrangians. This corresponds to an energy or derivative expansion in which operators are suppressed by a mass scale which is typically the mass of the 'integrated out' field. If the light fields present in the effective action are slowly varying, each term in the expansion is correspondingly smaller. Quantum mechanically, this corresponds to low energies. However, with non-local actions the curvature expansion has a different nature. Because non-local operators such as the inverse d' Alembertian $1/\nabla^2$ can appear, higher powers of the curvature such as $[(1/\nabla^2) R]^n$ are not automatically suppressed at low energy and the curvature expansion is not the same as the energy expansion. Instead, it is a way to describe the (calculable) infrared physics from quantum loops. The effects of these infrared non-local effects from loops are just starting to be explored \cite{Basem,Espriu,Cabrer,Maggiore,Woodard1,Woodard2,Woodard3,Woodard4,Barvreview,Calmet}.

In this paper we explore the non-local curvature expansion in a relatively simple setting - that of photons coupled to a massless charged scalar and to gravity. We also display the results relevant for massless fermions to highlight interesting features of the non-local action. Both the spacetime metric and the gauge field are treated as classical background fields. In a recent paper \cite{Donoghue}, we focused on obtaining the flat-space non-local effective action and the associated energy-momentum tensor that gives rise to the trace anomaly. Here we are concerned with generalizing the flat-space results to curved backgrounds. This is achieved via a technique that we refer to as the {\em non-linear completion} of the action where, similar to CPT, the action is displayed as an expansion in the curvatures. The non-local effective actions are a relatively unexplored topic and there remain interpretive issues that we explore here.

Most notable is the issue of the covariant nature of the non-local {\em form factors} such as $\ln \nabla^2$. In particular, we pay special attention to the generalization of the flat d' Alembertian to curved space which turns out to be a non-trivial aspect of the effective action. Moreover, direct use of the Feynman graph expansion of the effective action allows us to identify the terms which is related to the beta function of the theory and those which are not related to the latter. Our exploration leads to a better understanding of the non-local action that generates the QED trace (conformal) anomaly. To the best of our knowledge, this is an unsettled issue in the literature and the procedure of non-linear completion yields interesting insight into the correct form.

The plan of the paper is the following. In section \ref{boxproblem} we provide an overview of the main problem discussed in the paper and also present our results. In section \ref{remarks} we discuss some of the methodological issues with this program, pointing out the main difficulties of constructing non-local actions in curved spaces and in section \ref{nonlin} we describe the non-linear completion matching technique. Section \ref{sect3} is devoted for the non-linear completion of the quadratic action while the cubic action is displayed in section \ref{sect4}. We then move in section \ref{sect5} to show how the terms in the effective action generates the trace anomly. Finally, we summarize and conclude in section \ref{conc}.

\section{The problem of $\ln \Box$}\label{boxproblem}

In flat-space the one loop effective action for a photon, obtained by integrating out a massless charged scalar or fermion, has the form
\begin{equation}\label{quasilocal}
S = \int d^4x ~-\frac14 F_{\mu\nu} \left[\frac{1}{e^2(\mu)} - b_i \ln \left({\Box}/{\mu^2}\right)\right]F^{\mu\nu}
\end{equation}
where $b_i$ is the leading coefficient of the beta function, $b_s=1/(48\pi^2)$ for a charged scalar and $b_f=1/(12\pi^2)$ for a charged fermion, and $\Box = \partial^2$. Here the action is expressed in quasi-local form and the $\ln {\Box}/{\mu^2}$ operator is a shorthand for the fully non-local realization
\begin{equation}\label{posspacelog}
\langle x|\ln \left(\frac{\Box}{\mu^2}\right)|y\rangle \equiv L(x-y) = \int \frac{d^4q}{(2\pi)^4} e^{-iq\cdot (x-y)} \ln \left(\frac{-q^2}{\mu^2}\right)\ \ .
\end{equation}

When one desires a formulation in curved spacetime, one requires that the logarithm generalizes to the covariant form, with tensor indices raised and lowered with the metric, and the $\Box$ operator also being covariant. We will reserve the notation $\Box$ for the flat-space d'Alembertian and use $\nabla^2$ for the covariant version. That is, one requires
\begin{equation}\label{conformlog}
\frac{b_i}{4} \int d^4x ~ \eta^{\mu\alpha} \eta^{\nu\beta} F_{\mu\nu}  \ln \left({\Box}/{\mu^2}\right) F_{\alpha\beta} \to \frac{b_i}{4} \int d^4x \sqrt{-g} ~ g^{\mu\alpha}g^{\nu\beta} F_{\mu\nu}  \ln \left({\nabla^2}/{\mu^2}\right)F_{\alpha\beta}
\end{equation}
This can be made more usable through the definition of the log as
\begin{align}\label{massrep}
\ln \left({\nabla^2}/{\mu^2}\right) = - \int_0^\infty dm^2 \left[ \frac{1}{\nabla^2+m^2} - \frac{1}{\mu^2+m^2} \right] \ \,
\end{align}
which then involves propagators that can be covariantly defined. Even here the result is not simple as the inverse operator is acting on the tensor indices of $F_{\alpha\beta}$ and itself becomes a bitensor \cite{Poisson}. Later in the paper we expand the covariant form in eq. (\ref{conformlog}) to first order in the expansion $g_{\mu\nu}=\eta_{\mu\nu} +h_{\mu\nu}$, for photons satisfying $p^2=p'^2$, resulting in
\begin{align}\label{covariantlog}
\int d^4x \sqrt{g} \, F^{\alpha\beta} \ln \left( \frac{\nabla^2}{\mu^2} \right) F_{\alpha\beta} = \int d^4x \, \left[F^{\alpha\beta} \ln \left( \Box / \mu^2 \right) F_{\alpha\beta} + h_{\mu\nu} \left(\mathcal{O}_1^{\mu\nu} + \mathcal{O}_2^{\mu\nu}\right) \right]
\end{align}
where
\begin{align}\label{theOtensors}
\nonumber
&\mathcal{O}_1^{\mu\nu} = \frac{1}{2} \eta^{\mu\nu} F^{\alpha\beta} \ln(\Box/\mu^2) F_{\alpha\beta} - 2 F^{\mu}_{~\alpha} \log(\Box/\mu^2) F^{\nu\alpha}\\
&\mathcal{O}_2^{\mu\nu} = \partial^\mu \partial^\nu F_{\alpha\beta} \frac{1}{\Box} F^{\alpha\beta} + \partial^\mu \partial^\nu F^{\alpha\beta} \frac{1}{\Box} F_{\alpha\beta} - \eta^{\mu\nu} \partial_\lambda F_{\alpha\beta} \frac{1}{\Box} \partial^\lambda F^{\alpha\beta}
\end{align}
and indices are raised and lowered with the flat metric. We note that near the mass shell, $p^2=p'^2=\lambda^2\approx 0$, the $F(1/\Box)F$ terms are particularly dangerous as they involve the inverse photon ``mass''. Notice also that the logarithms in eq. (\ref{theOtensors}) are infrared singular.

On the other hand, in our previous work \cite{Donoghue}, we have explicitly calculated the $h_{\mu\nu}$ corrections to the effective action for a conformally coupled scalar field and on-shell photons, and have extracted the fermionic analogy from the work of \cite{Berends}. Interestingly, {\em none} of the above $h_{\mu\nu}$ terms in eq. (\ref{covariantlog}) are found in the result. Instead we get a relatively simple answer, in that the terms that are proportional to the beta function coefficient\footnote{There is also a term independent of the beta function which we include below.} are
\begin{equation}\label{calc}
\frac{b_i}{4}\Bigg\{F^{\alpha\beta}  \ln \left({\Box/\mu^2}\right)F_{\alpha\beta}  +  h_{\mu\nu} \left[2 \ln ({\Box}) T_{cl}^{\mu\nu} -  \frac23 \frac{1}{\Box} A^{\mu\nu} \right]\Bigg\}
\end{equation}
with
\begin{equation}\label{classicalT}
T^{cl}_{\mu\nu} = - F_{\mu\sigma} F_{\nu}^{~\sigma} + \frac{1}{4} g_{\mu\nu} F_{\alpha\beta}F^{\alpha\beta}
\end{equation}
and
\begin{align}
A_{\mu\nu} = \partial_\mu F_{\alpha\beta} \partial_\nu F^{\alpha\beta} + F_{\alpha\beta}\partial_\mu \partial_\nu F^{\alpha\beta} - \eta_{\mu\nu} \partial_\lambda F_{\alpha\beta} \partial^\lambda F^{\alpha\beta}
\end{align}
This result is itself generally covariant to this order in $h_{\mu\nu}$, although different in structure from eq. (\ref{covariantlog}). One can easily check that the {\em full} result is invarint under local coordinate transformations $h_{\mu\nu} \rightarrow h_{\mu\nu} + \partial_{(\mu} \xi_{\nu)}$. In contrast to eq. (\ref{covariantlog}) we see that eq. (\ref{calc}) does not contain any of the dangerous $F (1/\Box) F$ terms - the inverse photon mass does not arise in perturbation theory.

Both of the ${\cal O}(h_{\mu\nu})$ terms in eq. (\ref{calc}) are required by trace anomaly considerations and hence must be proportional to the
beta function coefficient $b_i$. The terms with the logarithm yield the correct trace anomaly for a pure scale transformation
\begin{equation}\label{purescaling}
x'=\lambda x, ~~A^\prime_\mu(x^\prime) = \lambda^{-1} A_\mu(x) ,~~ h'_{\mu\nu}(x')=h_{\mu\nu}(x), ~~\ln \Box' =  \ln \Box - \ln \lambda^2 ,
\end{equation}
where the non-invariance of $\ln \Box$
leads to
\begin{align}
T_{\mu}^{\, \mu} = \frac{b_i}{2} \left(\eta^{\mu\alpha} \eta^{\nu\beta} F_{\mu\nu} F_{\alpha\beta} + 2 h^{\mu\nu} T^{cl}_{\mu\nu} \right)
\end{align}
which is the correct expansion of the covariant density $\sqrt{g} F^2$. Under this rescaling the last term in eq. (\ref{calc}) is invariant. However, under a conformal transformation ($g_{\mu\nu} \rightarrow \exp{(2 \sigma(x))} g_{\mu\nu}$) restricted to flat-space
\begin{align}\label{Weylscaling}
h_{\mu\nu} \rightarrow h_{\mu\nu} + 2 \sigma \eta_{\mu\nu}
\end{align}
the first two terms in eq. (\ref{calc}) are invariant while the last term is not. Using the on-shell condition $\Box A_{\mu} = 0$ we have that\footnote{The details of these steps are carefully presented in \cite{Donoghue}.}
\begin{align}\label{trick}
\partial_{\lambda}F_{\alpha\beta}\partial^{\lambda}F^{\alpha\beta} = \frac{1}{2} \Box \left(F_{\mu\nu}F^{\mu\nu}\right),\quad \eta^{\mu\nu}\frac{1}{\Box}A_{\mu\nu}=-\frac32
F_{\mu\nu}F^{\mu\nu}
\end{align}
and we see that last term yields the correct trace anomaly. The two related transformations, rescaling the coordinates and rescaling the metric, act differently in the effective action yet both yielding the same anomaly relation. We see that both types of non-locality, i.e. the logarithm and the massless pole in eq. (\ref{calc}) are required by direct calculation as well as by anomaly considerations.

We seek the covariant curvature expansion which reproduces the perturbative results. For nomenclature, the term of order $F^2$ is referred to as second order in the curvature, while that with an extra gravitational curvature, e.g. $F^2R$, is called third order in the curvature. The details of the matching will be given in the body of the paper, while here we summarize the results.

The mismatch of the two expressions eqs. (\ref{covariantlog}) and (\ref{calc}) makes the expansion in the curvature relatively complicated. Because one is starting out with the $F\ln \nabla^2F$ expression as the covariant form which is second order in the curvature, one needs to add and subtract correction terms in order to reproduce the actual calculated result. These {\em counter-terms} are third order in the curvarure as we show below. This does not modify the covariance of the result - both expressions are covariant. Nevertheless it does make the resulting expression at third order quite complicated. This matching procedue, which we refer to as {\em non-linear completion}, occupies most of the work described below. We find that the result to this order in the curvature is\footnote{The placement of the differential operators appears somewhat different than the expressions in the body of the paper. This is allowed indeed under integration by parts as we are assuming asymptotically flat spacetimes.}
\begin{align}\label{generalizedlog}
\Gamma_{log}= &\frac{b_i}{4} \int d^4x \sqrt{g} \bigg\{ F_{\alpha\beta}  \ln \left({\nabla^2}/{\mu^2}\right)F^{\alpha\beta} -  \frac13 F_{\alpha\beta}F^{\alpha\beta}\frac{1}{\nabla^2} R  \nonumber \\
 &+4 R^{\mu\nu}\frac{1}{\nabla^2}\bigg[\log(\nabla^2)\left(-F_{\mu\sigma}F^{~\sigma}_\nu +\frac14 g_{\mu\nu}F_{\alpha\beta}F^{\alpha\beta}\right)\nonumber \\
 &+F_{\mu\sigma}\log(\nabla^2)F^{~\sigma}_\nu - \frac14 g_{\mu\nu}F_{\alpha\beta}\log(\nabla^2)F^{\alpha\beta}\bigg]\nonumber\\
 &+\frac13 R F_{\alpha\beta} \frac{1}{\nabla^2} F^{\alpha\beta} - C^\alpha_{~\beta\mu\nu} F_{\alpha}^{~\beta} \frac{1}{\nabla^2} F^{\mu\nu} \bigg\}
\end{align}
where $C^\alpha_{~\beta\mu\nu}$ is the Weyl tensor. Note that the logarithms within the square brackets $[...]$ do not need a factor of $\mu^2$ as the $\log \mu^2$ would cancel between the two terms. In particular, these terms are scale invariant as we will discuss later on.

We will show that eq. (\ref{generalizedlog}) has the correct anomaly properties. The way that this is accomplished is interesting. For a scale transformation as in eq. (\ref{purescaling}), it is the first term - the logarithm - which yields the anomaly. However for a local Weyl (conformal) transformation it is the second term - $F^2\frac{1}{\nabla^2} R$ - which is the active ingredient. This latter term appears as one of the portions of the Riegert anomaly action \cite{Riegert} when appropriately displayed in a curvature expansion. Finally for a global rescaling of the metric $g_{\mu\nu}\to e^{2\sigma}g_{\mu\nu}$, with $\sigma$ being a spacetime constant, there is a simpler path that again involves the logarithm. The latter is equivalent to a scale trasnformation as in eq. (\ref{purescaling}). We conclude that both the logarithm and the Riegert term (massless pole) are required by anomaly considerations. We comment on this dichotomy in regard to the geometric program to classify anomalies set forth by Deser and Schwimmer \cite{Schwimmer}.

Finally in order to match the full result found in the direct one-loop calculation \cite{Donoghue}, one must add a nonanomalous term that has no relation to the beta function
\begin{align}\label{Weylaction}
\Gamma_{Weyl} = n_C^i \int d^4x \, \sqrt{g} \, F^{\mu\nu} F_{\alpha}^{~\beta} \frac{1}{\nabla^2} C^{\alpha}_{~\beta\mu\nu}
\end{align}
where
\begin{align}
n_C^s =  - \frac{1}{96 \pi^2}, \quad n^f_C = \frac{1}{48 \pi^2} \ \ .
\end{align}
This is different for fermions and scalars and is invariant under both scaling and conformal transformations.

Our final result for the covariant one-loop effective action is
\begin{equation}
\Gamma_{tot} = S_{cl} +\Gamma_{log}+ \Gamma_{Weyl}
\label{total}
\end{equation}
where
\begin{equation}
S_{cl} = \int d^4x \sqrt{g} ~-\frac{1}{4e^2(\mu)} F_{\mu\nu}F^{\mu\nu}
\end{equation}
is the classical action.

\section{Covariant non-local actions: General remarks}\label{remarks}

General relativistic actions are readily described when local. Using the metric, covariant derivatives and curvature tensors one can construct
generally covariant local functions of the field variables. The ultraviolet divergences of quantum loops are therefore simple to treat because they are also
local \cite{tHooft,Nieu}. However non-local objects are difficult to describe in a generally covariant form because they sample the metric at a continuum of points in spacetime.
For a general metric, explicit expressions for such actions are not possible.

For massless scalar QED and after integrating out the charged scalars at one loop the effective action must be gauge invariant and thus involves only the field strength tensor. Up to quadratic order in the gauge field and using dimensional regularization, a general form in curved spacetime is
\begin{equation}\label{genaction}
\Gamma[g,A] = \frac{1}{e_0^2} S_{EM} + \int d^4x \int d^4y ~F_{\mu\nu}(x) M^{\mu\nu}_{\alpha\beta}(x,y;\mu) F^{\alpha\beta}(y)
\end{equation}
where $S_{EM}$ is the classical Maxwell action, $e_0$ is the bare electric charge and $M^{\mu\nu}_{\alpha\beta}(x,y;\mu)$ is an antisymmetric second-rank bi-tensor {\em density} of unit weight which explicitly depends on the renormalization scale. As we show below, this bi-tensor samples the full space-time and not just the pair of points $(x,y)$ since it involves the effects of massless
propagators. The practical question is what the form of this bi-tensor is and how we can best describe it.

The divergence contained in $M^{\mu\nu}_{\alpha\beta}(x,y)$ is local and calculable. It has the form \cite{Donoghue}
\begin{align}
M^{\mu\nu}_{\alpha\beta}(x,y;\mu)= \frac{1}{192 \pi^2 \epsilon} \sqrt{g(x)}^{1/4} \, \delta^4(x-y) \, \sqrt{g(y)}^{1/4} I^{\mu\nu}_{\alpha\beta} + L^{\mu\nu}_{\alpha\beta}(x,y;\mu)
\end{align}
where
\begin{align}
I^{\mu\nu}_{\alpha\beta} = \frac{1}{2} \left(\delta^{\mu}_{\alpha} \delta^{\nu}_{\beta} - \delta^{\mu}_{\beta} \delta^{\nu}_{\alpha} \right)
\end{align}
This divergence is absorbed into the renormalization of the electric charge. After removing this divergence, the residual bi-tensor $L^{\mu\nu}_{\alpha\beta}(x,y;\mu)$ is finite.

One might expect that there are also local terms proportional to the geometric curvatures, such as $FFR$ which would correspond to  $M^{\mu\nu,\alpha\beta}\sim \frac{g^{\mu\alpha}g^{\nu\beta} }{\sqrt{g}} \delta^4(x-y)~ R$. Such terms are found when one integrates out a {\em massive} charged particle \cite{Drummond}. However, they are absent in our problem, that of integrating out a massless field, simply on dimensional grounds. The curvatures involve two derivatives of the metric, and hence the coefficient of any local term of the form $FFR$ must have dimensions of $1/{\rm mass}^2$. Because all fields are massless, there is no way to obtain such a coefficient. Any factors of the curvature in the action must be balanced by non-local factors such as $1/\nabla^2$. This tells us that once we have dealt with charge renormalization, which is of course a local operator, the remainder of the effective
action will be purely non-local.

In flat space, the non-local function was obtained in \cite{Donoghue}
\begin{align}\label{flatNLfunc}
L^{(0) \mu\nu}_{~~ \alpha\beta}(x,y;\mu) = \frac{b_s e^2}{4} I^{\mu\nu}_{\alpha\beta} \, L(x-y;\mu)
\end{align}
where $b_s=1/(48\pi^2)$ is the leading coefficient of the QED beta function for a charged scalar, $e$ is the physical charge and $L(x-y;\mu)$ is displayed in eq. (\ref{posspacelog}).
As a warm-up for later usage, let us pause at this stage to show how one can convert from a non-local form to a quasi-local one employing non-local {\em form factors}. The latter are the building blocks of the curvature expansion.
Through the position-space representation
\begin{equation}\label{logbox}
\langle x|\ln \left(\frac{\partial^2}{\mu^2}\right)|y\rangle \equiv L(x-y)
\end{equation}
one can re-write eq. (\ref{genaction}) in quasi-local form as
\begin{equation}\label{flataction}
\Gamma^{(0)}[A] = S_{EM} + \frac{ b_s e^2}{4} \int d^4x ~ F_{\mu\nu} \left[\ln \left(\frac{{\partial^2}}{\mu^2}\right) \right] F^{\mu\nu} \ \ .
\end{equation}

To appreciate the subtleties in the construction of the bi-tensor, let us quote the effective action linear in metric perturbation around flat space $g_{\mu\nu} = \eta_{\mu\nu}+ h_{\mu\nu}$. In non-local form, it reads \cite{Donoghue}
\begin{align}
\Gamma^{(1)}[A,h]=-\frac12 \int d^4x \, \int d^4y \, h^{\mu\nu}(x) \left[ b_s \, L(x-y) T_{\mu\nu}^{cl}(y) - i \frac{b_s}{2} \, \Delta_F(x-y) \tilde{T}^{s}_{\mu\nu}(y) \right]
\end{align}
where photons are taken to be on-shell, i.e. dropping factors of $\Box F_{\mu\nu}$. Here we have defined
\begin{align}
\tilde{T}^s_{\mu\nu} = 2 \partial_{\mu}F_{\alpha\beta}\partial_{\nu}F^{\alpha\beta} - \eta_{\mu\nu}\partial_{\lambda}F_{\alpha\beta}\partial^{\lambda}F^{\alpha\beta}
\end{align}
and
\begin{align}\label{classicalemt}
T^{cl}_{\mu\nu} = - F_{\mu\sigma} F_{\nu}^{~\sigma} + \frac14 \eta_{\mu\nu} F_{\alpha\beta} F^{\alpha\beta}
\end{align}
is the classical energy-momentum tensor. We also have the massless propagator\footnote{The boundary condition imposed on the propagator depends on the application one is considering. For instance, for time-dependent systems one should choose the retarted propagator.}
\begin{align}
\Delta_F (x-y) = \int \frac{d^4 q}{(2\pi)^4} \frac{i}{q^2 + i0} e^{-i q \cdot(x-y)}  \ \ .
\end{align}
In this case, the result is local in the relative position of the gauge fields, but contains both logarithmic and massless-pole non-localities with respect
to the gravitational field. Allowing the gauge fields to be off-shell would lead to a non-locality in all three field variables due to the appearence of the triangle diagram\footnote{See the discussion in \cite{Donoghue}.}. Let us arrange the bi-tensor density at this order in metric perturbation, it reads
\begin{align}
L^{(1) \mu\nu}_{~~ \alpha\beta}(x,y;\mu) = \int d^4z \, h^{\sigma\lambda}(z) \, \left[ L(z-x;\mu) J^{\mu\nu}_{\alpha\beta\sigma\lambda} + \Delta_F(z-x) H^{\mu\nu}_{\alpha\beta\sigma\lambda} \right]  \delta^{(4)}(x-y)
\end{align}
where
\begin{align}
\nonumber
J^{\mu\nu}_{\alpha\beta\sigma\lambda} &= \frac{b_s}{8} \left(\delta^\nu_\alpha \delta^\mu_\sigma \eta_{\beta\lambda} + \delta^\nu_\alpha \delta^\mu_\lambda \eta_{\beta\sigma} - \delta^\mu_\alpha \delta^\nu_\sigma \eta_{\beta\lambda} - \delta^\mu_\alpha \delta^\nu_\lambda \eta_{\beta\sigma} - I^{\mu\nu}_{\alpha\beta} \eta_{\sigma\lambda}\right) \\
H^{\mu\nu}_{\alpha\beta\sigma\lambda} &= i \frac{b_s}{4} I^{\mu\nu}_{\alpha\beta} \left(2 \partial_\sigma \partial_\lambda - \eta_{\sigma\lambda} \Box \right) \ \ .
\end{align}
One immediately notices that the bi-tensor density samples the gravitational field over the whole spacetime manifold. This is the main reason that the explicit construction of such non-local objects is not possible in arbitrary geometries. Instead, one can use the quasi-local form factors to express the loop correction as follows
\begin{align}
\Gamma^{(1)}[A,h]=-\frac12 \int d^4x \, h^{\mu\nu}\left[ b_s \log \left(\frac{\Box}{\mu^2}\right)T_{\mu\nu}^{cl} + \frac{b_s}{2}\frac{1}{\Box} \tilde{T}^{s}_{\mu\nu} \right]
\end{align}
where the position-space representation of the inverse d' Alembertian is given above. In this paper, we seek a generally covariant non-linear completion of the above results that is accomplished by employing the non-local form factors.


\section{Non-linear completion: Expansion in the curvature}\label{nonlin}

The curvature expansion is a covariant method to display the effective action with arbitrary background fields. For local actions, the heat kernel expansion is the most elegant technique to resolve the functional determinant of any operator \cite{Avramidi1, Birrell, parkertoms, bos, Gilkey, dewitt}. Its usage encompases many applications in physics and mathematics, but unfortunately it becomes somewhat complicated when we deal with a massless operator. Moreover, the correspondance with the more familiar perturbative expansion of the effective action in terms of Feynman graphs is not very obvious \cite{Gorbar1,Gorbar2}. In this paper, we propose a new technique to obtain the effective action which we call non-linear completion. The logic is very similar to the matching procedure well known in effective field theory (EFT). This procedure proceeds by perturbative matching of the full theory onto the effective theory. What makes the construction of the EFT Lagrangian possible is the fact that it must inherent all the exact symmetries of the full theory. This is the pathway we are going to employ in our case as well.

In our example, the symmetries of the full theory are diffeomorphsim and gauge invariances and hence the non-local action must be constructed from the generalized curvartures. As we have shown in the previous section, the form factors are an important tool as they enable the action to be written in quasi-local form where the action is manifestly covariant. One starts by listing the relevant curvature basis and organize it in terms of a power series. For the example at hand, we have
\begin{align}\label{basis}
\nonumber
&\mathcal{R}^2: F_{\mu\nu}F^{\mu\nu} \\
&\mathcal{R}^3: F_{\mu\nu}F^{\mu\nu} R, \quad F_{\mu\alpha}F_\nu^{~\alpha} R^{\mu\nu}, \quad F^{\mu\nu} F^{\alpha\beta} R_{\mu\nu\alpha\beta}, \quad \nabla_\mu F^{\mu\nu} \nabla_\alpha F_\nu^{~\alpha} \ \ .
\end{align}
The field strength is the curvature of the gauge-connection and thus counts as one power of the curvature. The effective action will be displayed as an expansion in these {\em generalized} curvatures. The last operator in eq. (\ref{basis}) does not contribute when the photons are on-shell and thus we are not going to discuss it further. Then one proposes all possible non-local functionals of the d' Alembertian which could possibly act on the different terms in the curvarure basis
\begin{align}\label{ffactors}
\nonumber
&\mathcal{F}_2: \ln \left(\frac{\nabla^2}{\mu^2}\right) \\
&\mathcal{F}_3: \frac{1}{\nabla^2}, \quad \frac{\ln (\nabla^2_i/\mu^2) }{\nabla^2_i} \ \ .
\end{align}
where the subscripts denote the curvature upon which the operator acts. As far as $\mathcal{F}_3$ is concerned, one can arrange more operators such as
\begin{align}
\frac{\ln (\nabla^2_i / \nabla^2_j) }{f(\nabla^2)}
\end{align}
where $f(\nabla^2)$ is some function to be determined. However, we will see that no from factor of this kind arises in our example due to the on-shell condition. Although the above form factors look very complicated, these are all well defined via their Laplace transform
\begin{align}
\mathcal{F}(\nabla^2) = \int_0^\infty ds \, \mathcal{F}(s) e^{-s\nabla^2} \ \ .
\end{align}
The last step is perturbatively matching the full theory diagrams onto the non-local action. The 'Wilson' coefficients in this case only depends on the coupling constants of the full theory and are to be adjusted via the matching procedure. Since a massless field is being integrated out, these coefficients can not depend on any mass or renormalization scale, i.e. the non-local action is completely insensitive to the UV.


\section{The $\mathcal{R}^2$ action: The elusive logarithm}\label{sect3}

In this section, we discuss the non-linear completion of the flat-space action in eq. (\ref{flataction}). It reads
\begin{align}\label{quadaction}
\overset{\scriptscriptstyle{(2)}}{\Gamma}[g,A] = \frac{b_s}{4} \int d^4x \sqrt{g}\, g^{\mu\alpha} g^{\nu\beta} \, F_{\alpha\beta} \log\left(\frac{\nabla^2}{\mu^2}\right) F_{\mu\nu}
\end{align}
where $\nabla^2 = g^{\mu\nu} \nabla_\mu \nabla_\nu$ is the covariant d' Alembertian. The matching onto eq. (\ref{flataction}) is immediate. Now one must raise the question: what is the expansion of the above action around flat space? In partiuclar, the piece linear in the metric perturbation and its connection to the perturbative computation. The answer to these questions is very important in understanding the covariant nature of the quasi-local expansion. In the remainder of this section, we show how to consistently expand the logarthim and prove that the $\mathcal{O}(h)$ term in the action is entirely absent from the perturbative computation. We start by showing the steps for a scalar field as a toy example and then discuss the more interesting example of a 2-form.

\subsection{Toy example: A scalar field}

Let us consider the following action
\begin{align}
\Gamma[g,\phi] = \int \, d^4x \, \sqrt{g} \, \phi \ln\left(\frac{\nabla^2}{\mu^2}\right) \phi \ \ .
\end{align}
The goal is to expand the action around flat space to linear order in the metric perturbation $g_{\mu\nu} = \eta_{\mu\nu} + h_{\mu\nu}$. The most convenient way to accomplish this is to first vary the action with respect to the metric and then restrict the result to flat space. Using eq. (\ref{massrep}), we find
\begin{align}\label{actionvar}
\delta_g \Gamma[\eta,\phi] = \int d^4x \int_0^{\infty} dm^2 \, \big\{ \phi \, \left(\Box + m^2\right)^{-1} [\delta_g \nabla^2]_{g=\eta} \left(\Box + m^2\right)^{-1}  \, \phi \big\}+ ... \ \ .
\end{align}
where the ellipses denote terms resulting from the variation of $\sqrt{g}$ which do not matter to our discussion. To arrive at the above expression, we have used the formal variation of an inverse operator
\begin{align}
\delta_g \frac{1}{\nabla^2 + m^2} = - \frac{1}{\nabla^2 + m^2} (\delta_g \nabla^2) \frac{1}{\nabla^2 + m^2} \ \ .
\end{align}
The variation of the d' Alembertian depends on the tensor field in the action. For a scalar field, we have
\begin{align}\label{nablaonscalar}
(\delta_g \nabla^2) \Psi = \left(\delta g^{\mu\nu} \partial_\mu \partial_\nu - \delta g^{\mu\nu} \Gamma^\alpha_{\mu\nu} \partial_\alpha - g^{\mu\nu} \delta\Gamma^\alpha_{\mu\nu} \partial_\alpha \right) \Psi
\end{align}
where
\begin{align}
\delta\Gamma^\alpha_{\mu\nu} = \frac{1}{2} g^{\alpha\beta} \left(\partial_\mu \delta g_{\beta\nu} + \partial_\nu \delta g_{\beta\mu} - \partial_\beta \delta g_{\mu\nu} \right) \ \ .
\end{align}
It is advisable at this stage to express eq. (\ref{actionvar}) in a non-local form which is accompliahed via the identity
\begin{align}\label{propid}
\frac{1}{\Box + m^2} \Psi = \int d^4y \, \Delta(x-y) \Psi(y)
\end{align}
where
\begin{align}
(\Box + m^2) \Delta(x-y) = \delta^{(4)}(x-y) \ \ .
\end{align}
If we recall that $\delta g^{\mu\nu} = - h^{\mu\nu}$ around flat space, we find
\begin{align}
\nonumber
\delta_g \Gamma[\eta,\phi] &= \int d^4x d^4y d^4z \int_0^\infty dm^2 \, \phi(x) \Delta(x-y) \\
&\left(-h^{\mu\nu} \partial_\mu \partial_\nu - \partial^\mu h_{\mu\nu} \partial^\nu + \frac{1}{2} \partial^\alpha h \partial_\alpha \right) \Delta(y-z) \phi(z)
\end{align}
where
\begin{align}
\Delta(x-y) = - \int \frac{d^4 l}{(2\pi)^4} \frac{e^{-il \cdot (x-y)}}{l^2 - m^2} \ \ .
\end{align}
Although the above must be defined with some boundary condition, this is not going to affect our discussion.  Notice that one could obtain the same result using the more explicit variation of the propagator
\begin{align}
\frac{\delta G(x,x^\prime)}{\delta g^{\mu\nu}(z)} = - \int\, d^4y \, G(x,y) \left[\frac{\delta \nabla^2}{\delta g^{\mu\nu}(z)}\right] G(y,x^\prime) \ \ .
\end{align}
To facilitate comparison with the perturbative calculation, we can Fourier transform the above expression and find
\begin{align}
\int \frac{d^4p}{(2\pi)^4} \frac{d^4p^\prime}{(2\pi)^4} \phi(p) \phi(p^\prime) h^{\mu\nu}(q) \mathcal{P}_{\mu\nu} \frac{\ln p^{\prime 2} - \ln p^2}{p^2 - p^{\prime 2}}, \quad q = -p-p^\prime
\end{align}
where
\begin{align}
\mathcal{P}_{\mu\nu} = \frac{1}{2} p^\prime_\mu p^\prime_\nu + \frac{1}{2} p_\mu p_\nu + \frac{1}{4} q_\mu p^\prime_\nu + \frac{1}{4} q_\nu p^\prime_\mu +\frac{1}{4} q_\mu p_\nu + \frac{1}{4} q_\nu p_\mu - \frac{1}{4} q \cdot p^\prime \eta_{\mu\nu} - \frac{1}{4} q \cdot p \eta_{\mu\nu}
\end{align}

\subsection{2-forms}

We now turn to the treatment of 2-forms which is our main interest. There are two distinct pieces that arise from the variation procedure. The first comes from varying the explicit factors of the metric tensor in eq. (\ref{quadaction}) while the second comes from varying the logarithm and the procedure is almost identical to the scalar example aside from some differences related to the tensor rank that we now discuss. First, we generalize eq. (\ref{nablaonscalar}) to the variation of the d' Alembertian when it acts on a 2-form

\begin{align}\label{nablaontensor}
\nonumber
(\delta_g \nabla^2 A_{\mu\nu})|_{g=\eta} = &\left(-h^{\alpha\beta} \partial_\alpha \partial_\beta - \eta^{\alpha\beta} \delta\Gamma^\sigma_{\alpha\beta} \partial_\sigma \right) A_{\mu\nu} - \partial^\beta \left(\delta\Gamma^\sigma_{\beta\mu} A_{\sigma\nu} + \delta\Gamma^\sigma_{\beta\nu} A_{\sigma\mu}  \right)\\
&-\delta\Gamma^\sigma_{\beta\mu} \partial^\beta A_{\sigma\nu} - \delta\Gamma^\sigma_{\beta\nu} \partial^\beta A_{\mu\sigma} \ \ .
\end{align}
Second, we need to generalize eq. (\ref{propid})
\begin{align}\label{propidtensor}
\frac{1}{\Box+m^2} A_{\mu\nu} = \int d^4y \, \Delta_{\mu\nu}^{\alpha\beta}(x-y) A_{\alpha\beta}(y), \quad \Delta_{\mu\nu}^{\alpha\beta} = I_{\mu\nu}^{\alpha\beta} \Delta(x-y) \ \ .
\end{align}
We recognize in eq. (\ref{nablaontensor}) a structure identical to the scalar field and the result is the same as before but with the difference that both transversality and on-shellness are taken into account as described in eq. (\ref{transcond}). We now show how to treat the new structures in eq. (\ref{nablaontensor}). In position-space, we have the following piece
\begin{align}
\nonumber
\int d^4x d^4y d^4z \int_0^\infty dm^2 &A^{\mu\nu}(x) \Delta(x-y) \bigg[-2\delta\Gamma^\sigma_{\lambda\mu}(y) (\partial^\lambda \Delta(y-x)) A_{\sigma\nu}(z) \\
&- (\partial^\lambda \delta\Gamma^\sigma_{\lambda\mu}(y)) \Delta(y-z) A_{\sigma\nu}(z) \bigg]
\end{align}
where we used eq. (\ref{propidtensor}). We now have all ingredients and after a laborious computation in momentum-space one finds
\begin{align}\label{linearlog}
\overset{\scriptscriptstyle{(2)}}\Gamma[g,A] = \Gamma^{(0)}[A] + \frac{b_s}{4} \int \frac{d^4p}{(2\pi)^4} \frac{d^4p^\prime}{(2\pi)^4} A^\alpha(p) A^\beta(-p^\prime) h^{\mu\nu}(-q) \, \left(\mathcal{D}_{\alpha\beta\mu\nu} - \mathcal{N}_{\alpha\beta\mu\nu}\right) + \mathcal{O}(h^2)
\end{align}
where
\begin{align}
\mathcal{D}_{\mu\nu\alpha\beta} &= \frac{1}{2} (q^2 \eta_{\mu\nu} - q_\mu q_\nu - Q_\mu Q_\nu )(p \cdot p^\prime \eta_{\alpha\beta} - p_\alpha^\prime p_\beta) \frac{\ln p^{\prime 2} - \ln p^2}{p^2 - p^{\prime 2}}\\
\mathcal{N}_{\mu\nu\alpha\beta} &= \mathcal{M}^0_{\mu\nu\alpha\beta} \log\left(\frac{-p^2}{\mu^2}\right)
\end{align}
and $\mathcal{M}^0_{\mu\nu,\alpha\beta}$ is
the tensor is the lowest-order matrix element describing the local coupling of photons to gravity. Explicitly, it reads
\begin{align}\label{lowestorder}
\nonumber
\mathcal{M}^0_{\mu\nu,\alpha\beta}&=p_{\mu}^{\prime} p_{\nu} \eta_{\alpha\beta} + p_{\mu} p^{\prime}_{\nu} \eta_{\alpha\beta} + \eta_{\mu\nu} p^{\prime}_{\alpha} p_{\beta} - p_{\mu} p^{\prime}_{\alpha} \eta_{\nu\beta} - p_{\mu}^{\prime} p_{\beta} \eta_{\alpha\nu} - p_{\nu} p^{\prime}_{\alpha} \eta_{\mu\beta} \\
&- p_{\nu}^{\prime} p_{\beta} \eta_{\alpha\mu} + p \cdot p^{\prime} (\eta_{\mu\alpha} \eta_{\beta\nu} + \eta_{\mu\beta} \eta_{\nu\alpha} - \eta_{\mu\nu} \eta_{\alpha\beta}) \ \ .
\end{align}

The first tensor is the result of varying the metric tensor inside the logarithm, while the second comes from the metric tensors in the rest of the action. Notice that we enforce both transversality and on-shellness except in non-analytic expressions that are infrared singular. Apart from being gauge-invariant, the above tensors respects local energy-momentum conservation
\begin{align}
\nonumber
&q^\mu \mathcal{D}_{\mu\nu\alpha\beta} = q^\nu \mathcal{D}_{\mu\nu\alpha\beta} = 0 \\
&q^\mu \mathcal{N}_{\mu\nu\alpha\beta} = q^\nu \mathcal{N}_{\mu\nu\alpha\beta} = 0 \ \ .
\end{align}
Indeed this property is guaranteed for the tensor $\mathcal{N}_{\mu\nu\alpha\beta}$ since it is the variation of a local operator, but it is gratifying to see that the same applies for $\mathcal{D}_{\mu\nu\alpha\beta}$ which is the variation of a purely non-local object.


\section{The $\mathcal{R}^3$ action}\label{sect4}

In this section, we perform the matching procedure outlined in section \ref{nonlin}. It is more convenient to work in momentum space, and so we list the momentum-space expansions of the different curvature invariants in an appendix.

\subsection{Terms including $1/\nabla^2$}

Here we display the non-linear completion of the anomalous contribution to the effective action. At the linear level, we had \cite{Donoghue}
\begin{align}
\Gamma_{pole}[A,h] = \int \frac{d^4p}{(2\pi)^4} \frac{d^4p^\prime}{(2\pi)^4} A^\alpha(p) A^\beta(-p^\prime) h^{\mu\nu}(-q) \frac{1}{q^2} \mathcal{M}^s_{\mu\nu\alpha\beta}
\end{align}
where
\begin{align}\label{flatanomaly}
\mathcal{M}^s_{\mu\nu\alpha\beta} = \frac{1}{192\pi^2} (p_\alpha^\prime p_\beta - p \cdot p^\prime \eta_{\alpha\beta}) (Q_\mu Q_\nu - q_\mu q_\nu + q^2 \eta_{\mu\nu}) \ \ .
\end{align}
The non-linear completion commences by proposing the ansatz
\begin{align}\label{1overbox}
\nonumber
\overset{\scriptscriptstyle{(3)}}{\Gamma}_{pole}[g,A] = \int d^4x \sqrt{g} \, \bigg( &P^S F_{\mu\nu} F^{\mu\nu} \frac{1}{\nabla^2} R + P^{Ric} F^\beta_{~\mu} F^{\alpha\mu} \frac{1}{\nabla^2} R_{\alpha\beta} \\
&+ P^{Riem} F_{\alpha}^{~\beta} F^{\mu\nu} \frac{1}{\nabla^2} R^\alpha_{~\beta\mu\nu} \bigg)
\end{align}
where the choice of the form factor is easily motivated by the presence of the massless pole
\begin{align}
\frac{1}{-q^2} \rightarrow \frac{1}{\Box} \ \ .
\end{align}
Using the expansions provided in the appendix, one can form a linear system to solve for the three coefficients. It naively appears that the system is overdetermined since the expansion of the curvature invariants contain tensor structures that do not appear in eq. (\ref{flatanomaly}). Nevertheless, one only finds exactly three {\em independent} equations which uniquely yields
\begin{align}
P^S = -\frac{1}{192\pi^2}, \quad P^{Ric} = \frac{1}{48\pi^2}, \quad P^{Riem} = - \frac{1}{96\pi^2} \ \ .
\end{align}
We can use the Weyl tensor to change the curvature basis which is very useful to discuss the conformal (non)-invariance of the effective action. In 4$D$, the Weyl tensor reads
\begin{align}\label{weyl}
C_{\mu\nu\alpha\beta} = R_{\mu\nu\alpha\beta} - \frac12 \big(g_{\mu\alpha} R_{\nu\beta} - g_{\mu\beta} R_{\nu\alpha} - g_{\nu\alpha} R_{\mu\beta} + g_{\nu\beta} R_{\mu\alpha} \big) + \frac{R}{6} \big(g_{\mu\alpha} g_{\nu\beta} - g_{\mu\beta} g_{\nu\alpha}\big) \ \ .
\end{align}
Hence, eq. (\ref{1overbox}) becomes
\begin{align}\label{1overboxweyl}
\overset{\scriptscriptstyle{(3)}}{\Gamma}_{pole}[g,A] = \int d^4x \sqrt{g} \big(\bar{P}^S F_{\mu\nu} F^{\mu\nu} \frac{1}{\nabla^2} R + P^{C} F_{\alpha}^{~\beta} F^{\mu\nu} \frac{1}{\nabla^2} C^\alpha_{~\beta\mu\nu} \big)
\end{align}
where
\begin{align}
\bar{P}^S = -\frac{1}{576\pi^2}, \quad P^C = -\frac{1}{96\pi^2} \ \ .
\end{align}
In fact, the coefficient of the Ricci scalar piece is indeed related to the beta function of the theory as could easily be checked by consulting the effective action in fermionic QED \cite{Donoghue}. One finds
\begin{align}
\bar{P}^S = -\frac{b_i}{12}, \quad b_{boson} = \frac{1}{48\pi^2}, \quad b_{fermion} = \frac{1}{12\pi^2} \ \ .
\end{align}

\subsection{Terms including $(\log \nabla^2)/\nabla^2$}

In the linear action, we also found a logarithmic non-locality which reads \cite{Donoghue}
\begin{align}
\Gamma[A,h] = -\frac{b_i}{4} \int \frac{d^4p}{(2\pi)^4} \frac{d^4p^\prime}{(2\pi)^4} A^\alpha(p) A^\beta(-p^\prime) h^{\mu\nu}(-q) \log\left(\frac{-q^2}{\mu^2}\right) \mathcal{M}^0_{\mu\nu\alpha\beta}
\end{align}
where $\mathcal{M}^0_{\mu\nu,\alpha\beta}$ has been given previously in eq. (\ref{lowestorder}).
Although the appearance of $\mathcal{M}^0$ might suggest that the above action could be matched onto the quadratic basis, this is in fact impossible. We show next that the action can only be matched onto the cubic basis with the following form factor
\begin{align}\label{logoverbox}
\nonumber
\overset{\scriptscriptstyle{(3)}}{\Gamma}_{log}[g,A] = \int d^4x \sqrt{g} \, \bigg(&L^S F_{\mu\nu} F^{\mu\nu} \frac{\log(\nabla^2/\mu^2)}{\nabla^2} R + L^{Ric} F^\beta_{~\mu} F^{\alpha\mu} \frac{\log(\nabla^2/\mu^2)}{\nabla^2} R_{\alpha\beta} \\
&+ L^{Riem} F_{\alpha}^{~\beta} F^{\mu\nu} \frac{\log(\nabla^2/\mu^2)}{\nabla^2} R^\alpha_{~\beta\mu\nu} \bigg) \ \ .
\end{align}
The $1/\Box$ is inserted for dimensional consistency at this stage as it comprises the only possible non-local object one can employ. The matching procedure is the only way to decide on the consistency of the ansatz. Once again, using the curvature expansions in the appendix one ends up with three {\em independent} equations which uniquely fixes the coefficients
\begin{align}
L^S = \frac{b_s}{4}, \quad L^{Ric} = - b_s, \quad L^{Riem} = 0 \ \ .
\end{align}
The $1/q^2$ factor which results from inserting the inverse d' Alembertian cancels out against factors of $q^2$ in the curvarure invariants. Using eq. (\ref{classicalemt}), one can rewrite the above action in a more transparent form which will prove useful in discussing the conformal (non)-invariance of the action
\begin{align}\label{logboxoverbox}
\overset{\scriptscriptstyle{(3)}}{\Gamma}_{log}[g,A] = b_s \int d^4x \sqrt{g}\, T^{cl}_{\mu\nu} \frac{\log(\nabla^2/\mu^2)}{\nabla^2} R^{\mu\nu}
\end{align}

\subsection{Counterterms for the logarithm}

Here we display the counterterms that we need to cancel out the $\mathcal{O}(h)$ piece that appears in the expansion of the quadratic action eq. (\ref{linearlog}). As we show next, these are third order in the curvature. There are two independent tensors in eq. (\ref{linearlog}) which should be matched onto two different ansatz. For the tensor $\mathcal{N}_{\mu\nu\alpha\beta}$, the ansatz is the following
\begin{align}\label{counter1}
\nonumber
\overset{\scriptscriptstyle{(3)}}{\Gamma}_{ct.1}[g,A] = \int d^4x \sqrt{g}\, \bigg[&C^S F_{\mu\nu} \log(\nabla^2/\mu^2) F^{\mu\nu} \frac{1}{\nabla^2} R + C^{Ric} F^\beta_{~\mu} \log(\nabla^2/\mu^2) F^{\alpha\mu} \frac{1}{\nabla^2} R_{\alpha\beta} \\
&+ C^{Riem} F_{\alpha}^{~\beta} \log(\nabla^2/\mu^2) F^{\mu\nu} \frac{1}{\nabla^2} R^\alpha_{~\beta\mu\nu} \bigg] \ \ .
\end{align}
A straightforward matching as before yields
\begin{align}
C^S = -\frac{b_s}{4}, \quad C^{Ric} = b_s, \quad C^{Riem} = 0 \ \ .
\end{align}
Moving to the tensor $\mathcal{D}_{\mu\nu\alpha\beta}$, we first notice that in the limit $p^2 = p^{\prime 2}$ the non-analytic structure becomes
\begin{align}
\lim_{ p^{\prime 2} \rightarrow p^2} \frac{\ln p^{\prime 2} - \ln p^2}{p^2 - p^{\prime 2}} = - \frac{1}{p^2}
\end{align}
which enables us to propose the following ansatz
\begin{align}\label{counter2}
\nonumber
\overset{\scriptscriptstyle{(3)}}{\Gamma}_{ct.2}[g,A] = \int d^4x \sqrt{g}\, \bigg[&T^S F_{\mu\nu} \frac{1}{\nabla^2} F^{\mu\nu}  R + T^{Ric} F^\beta_{~\mu} \frac{1}{\nabla^2}  F^{\alpha\mu} R_{\alpha\beta}\\
&+ T^{C} F_{\alpha}^{~\beta} \frac{1}{\nabla^2} F^{\mu\nu} C^\alpha_{~\beta\mu\nu} \bigg] \ \ .
\end{align}
We choose to work directly in the conformal basis, since it is more convenient. The matching yields
\begin{align}
T^S = \frac{b_s}{12}, \quad T^{Ric} = 0, \quad T^C = - \frac{b_s}{4} \ \ .
\end{align}
The same result holds for fermions, substituting $b_f$ for $b_s$.


\section{Remarks on the trace anomaly}\label{sect5}

In this section we explore the conformal transformation properties of the different terms in the action\footnote{See a parallel discussion in \cite{Zhytnikov}.}. We find an interesting dichotomy regarding the terms that give rise to the anomaly in response to conformal transformations. This requires a separate treatment of scale (global) and Weyl (local) transformations. Since the seminal work of Deser, Duff and Isham \cite{Duff}, there has been a consistent effort to understand the precise form of the non-local effective action that gives rise to gravitational anomalies.
In \cite{Schwimmer}, anomalies were geometrically classified to fall into two types. Type A anomalies arise from scale-invariant actions, i.e. invariant under a global Weyl rescaling. These are unique and strictly proportional to the Euler density of the dimension. On the other hand, type B anomalies arise from scale-dependent actions\footnote{This also means that the action carries an explicit dependence on the renormalization scale $\mu$.} but the local anomaly itself when {\em denstized} is invariant under local Weyl tranformations. For example, for a massless minimally coupled scalar in 2D the anomaly reads
\begin{align}
T_\mu^{~\mu} = \frac{1}{24\pi} R
\end{align}
whose density $\sqrt{g} R$ is indeed the Euler density in 2D. So this is a type A anomaly, and one can check easily that the non-local Polyakov action \cite{Polyakov} giving rise to the anomaly is scale-invariant. Reigert, following Polyakov, constructed a non-local action in 4D by integrating the anomaly \cite{Riegert}. However, the Riegert action was criticized in \cite{Schwimmer,Deser1,Erdmenger} based on several reasons while others \cite{Giannotti,Mazur} argued for its validity.

The QED trace anomaly falls into type B since its denstized version is indeed (locally) conformally invariant, and according to the above classification the generating non-local action should be scale-dependent. We show below that the two non-local structures present in the action are required to generate the correct trace relation whether one performs a global or local conformal transformation. Remarkably, the different terms have completely different behavior under both types of transformations. In particular, the trace relation is generated from the logarithmic non-locality under a scale transformation while the massless pole non-locality is responsible for the latter under local ones.

\subsection{Weyl transformations}

Let us commence by considering local transformations. Under an infinitismal transformation, we have
\begin{align}
\delta_\sigma g_{\mu\nu} = 2 \sigma(x) g_{\mu\nu}
\end{align}
which leads to the following transformation of the Christoffel symbol
\begin{align}
\delta_\sigma \Gamma^\lambda_{\mu\nu} = \delta^\lambda_\mu \nabla_\nu \sigma + \delta^\lambda_\nu \nabla_\mu \sigma - g_{\mu\nu} \nabla^\lambda \sigma \ \ .
\end{align}
From these one readily determines the tranformation of the different curvature tensors. The ones we need are
\begin{align}\label{curvtrans}
\delta_\sigma R_{\mu\nu} = 2 \nabla_\mu \nabla_\nu \sigma + g_{\mu\nu} \nabla^2 \sigma, \quad \delta_\sigma R = 6 \nabla^2 \sigma - 2 \sigma R \ \ .
\end{align}

Another object we will need its transformation is the d' Alembertian operator acting on different tensors, in particular, 2-forms
\begin{align}\label{sigma2form}
\delta_\sigma (\nabla^2 A_{\mu\nu}) = -2 \sigma \nabla^2 A_{\mu\nu} - 2 (\nabla^2 \sigma) A_{\mu\nu} - 2 (\nabla_\mu \sigma) \nabla^{\lambda} A_{\lambda\nu}
+2 (\nabla_\nu \sigma) \nabla^{\lambda} A_{\mu\lambda} \ \ .
\end{align}
where it is understood that $A_{\mu\nu}$ is invariant. Once again, let us apply the transformation to the quadratic action
\begin{align}
\delta_\sigma \overset{\scriptscriptstyle{(2)}}{\Gamma}[e^{2\sigma}g,A] = \frac{b_s}{4} \int d^4x \sqrt{g} \int_0^\infty dm^2 \, F^{\mu\nu} (\nabla^2 + m^2)^{-1} \, (\delta_\sigma \nabla^2) (\nabla^2 + m^2)^{-1} F_{\mu\nu} \ \ .
\end{align}
Counting the powers of curvature is very important at this stage. The function $\sigma(x)$ counts as a power of the curvature which means that we can freely commute covariant derivatives. For example,
\begin{align}\label{commute}
[\nabla_\mu , (\nabla^2 + m^2)^{-1} ] \sim \mathcal{O}(R) \ \ .
\end{align}
Using eq. (\ref{sigma2form}) and integrating by parts, we find
\begin{align}\label{sigvar2form}
\nonumber
\delta_\sigma \overset{\scriptscriptstyle{(2)}}{\Gamma}[e^{2\sigma}g,A] = \frac{b_s}{4} &\int d^4x \sqrt{g} \int_0^\infty dm^2 \,  F^{\mu\nu} (\nabla^2 + m^2)^{-1} (\nabla^2 + m^2)^{-1} \\
&\left(-2 \sigma \nabla^2 F_{\mu\nu} - 2 (\nabla^2 \sigma) F_{\mu\nu} + 2 \sigma \nabla^\lambda \nabla_\mu F_{\lambda\nu}
-2 \sigma \nabla^\lambda \nabla_\nu F_{\mu\lambda}\right) \ \ .
\end{align}
Now we employ the Bianchi identity
\begin{align}
\nabla_\mu F_{\lambda\nu} + \nabla_\nu F_{\mu \lambda} + \nabla_\lambda F_{\nu\mu} = 0
\end{align}
to find
\begin{align}\label{sigvarfinal}
\delta_\sigma \overset{\scriptscriptstyle{(2)}}{\Gamma}[e^{2\sigma}g,A] =  - \frac{b_s}{2} &\int d^4x \sqrt{g} \, (\nabla^2 \sigma) F^{\mu\nu} \frac{1}{\nabla^2} F_{\mu\nu} \ \ .
\end{align}
Although a prescription to integrate over ($dm^2$) might not seem obvious with the inverse operators present in eq. (\ref{sigvar2form}), one could easily check the above equation by linearlizing eq. (\ref{sigvar2form}) around flat space. It is very important to notice that the above computation clearly shows that under the local transformation the log piece does not give rise to the anomaly as it does not possess the correct pole structure. Moreover, we show next that eq. (\ref{sigvarfinal}) cancels identocally against the contribution coming from the transformation of the counter-term.

Indeed we need not worry about terms containing the Weyl tensor. Moreover, from the transformation listed in eq. (\ref{curvtrans}) one easily finds
\begin{align}
\delta_\sigma \overset{\scriptscriptstyle{(3)}}{\Gamma}_{log}[e^{2\sigma}g,A] = \delta_\sigma \overset{\scriptscriptstyle{(3)}}{\Gamma}_{ct.1}[e^{2\sigma}g,A] = 0
\end{align}
given that the field strength is on-shell
\begin{align}
\nabla_\mu F^{\mu\nu} = 0 \ \ .
\end{align}
The other counter-term transforms as
\begin{align}
\delta_\sigma \overset{\scriptscriptstyle{(3)}}{\Gamma}_{ct.2}[e^{2\sigma}g,A] = \frac{b_s}{2} &\int d^4x \sqrt{g} \, (\nabla^2 \sigma) F^{\mu\nu} \frac{1}{\nabla^2} F_{\mu\nu}
\end{align}
exactly cancelling eq. (\ref{sigvarfinal}) as promised.

Lastly, the massless pole non-locality of eq. (\ref{1overboxweyl}) is the piece that yields the correct trace. To this order in the curvature we only need to keep the $\delta_\sigma R= 6 \nabla^2 \sigma +...$ term in the transformation of eq. (\ref{curvtrans}), and neglect the variation of $1/\nabla^2$. Doing this yields
\begin{align}\label{poletrans2}
\delta_\sigma \overset{\scriptscriptstyle{(3)}}{\Gamma}_{pole}[e^{2\sigma}g,A] = - \frac{b_s}{2} &\int d^4x \sqrt{g} \, \sigma F^{\mu\nu} F_{\mu\nu} \ \ .
\end{align}
which yields the desired trace. In order to see this more simply, and make contact with the literature, we can show that all corrections to this result are higher order in the curvature by employing the Riegert action \cite{Riegert}. By defining the Paneitz operator \cite{Erdmenger2}
\begin{align}\label{Paneitz}
\Delta_4 = \nabla^2\nabla^2 +2\nabla_\mu(R^{\mu\nu} -\frac13 g^{\mu\nu}R)\nabla_\nu
\end{align}
and
\begin{align}
{\cal R}=\sqrt{-g}\left(\nabla^2 R- \frac32 G\right)
\end{align}
where $G$ is the Gauss-Bonnet term
\begin{align}
G =R^{\alpha\beta\gamma\delta}R_{\alpha\beta\gamma\delta} - 4R^{\alpha\beta}R_{\alpha\beta}+R^2
\end{align}
we can see that the Riegert form of this action
\begin{align}\label{Riegert}
{\Gamma}_{R}[g,A] = \int d^4x \sqrt{g} \big(\bar{P}^S F_{\mu\nu} F^{\mu\nu} \frac{1}{\Delta_4}{\cal R}  \big)
\end{align}
is equivalent to the first term of eq. (\ref{1overboxweyl}) up to terms which are higher order in the curvature, ${\Gamma}_{R}[g,A] = \overset{\scriptscriptstyle{(3)}}{\Gamma}_{pole}[g,A] + {\cal O}(F^2R^2)$. With this form, one can show without
approximation \cite{Erdmenger2} that
\begin{align}
\delta_\sigma \frac{1}{\Delta_4}=0~~, ~~~~~~~~~~\delta_\sigma {\cal R}=6 \Delta_4 \sigma
\end{align}
yielding
\begin{align}\label{poletrans}
\delta_\sigma {\Gamma}_{R}[g,A] = - \frac{b_s}{2} &\int d^4x \sqrt{g} \, \sigma F^{\mu\nu} F_{\mu\nu} \ \ .
\end{align}
The expansion in the curvature has yielded a term which, to this order in the curvature, is equivalent to the Reigert action.

Now we know that a conformal variation of a generic action reads
\begin{align}\label{traceid}
\delta_{\sigma} S = - \int d^4x \sqrt{g} \, \sigma \, T_\mu^{~\mu}
\end{align}
and thus indeed eq. (\ref{poletrans}) (likewise eq. (\ref{poletrans2})) yields the correct trace relation.

\subsection{Scale transformations}

A global scale transformation can take a couple of forms. One involves the scaling relations shown in eq. (\ref{purescaling}). It is simple to see that this transformation leaves all terms invariant, except the covariant logarithm. The logarithmic terms inside the square brackets [...] of eq. (\ref{generalizedlog}) are both shifted by $\ln \nabla^{2} \to  \ln \nabla^2 - \ln \lambda^2$, but $\ln \lambda^2$ cancels out leaving the whole expression invariant. So in contrast to the above Weyl transformation, this form of rescaling yields an anomaly that comes from the covariant logarithm.

Interestingly in the presence of the metric, there is another way to achieve a global scale transformation. In this case the transformation on the metric acts as follows
\begin{align}
g_{\mu\nu} \rightarrow e^{2\sigma} g_{\mu\nu}
\end{align}
where $\sigma$ is a constant, {\em not} necessarily infinitesimal. This may seem like a subcase of the Weyl transformation, but in fact it is distinct \cite{Polchinski}. Computationally, a distinction arises in that derivatives of $\sigma$ vanish, so that many of the integration-by-parts steps from the previous section are not available.

In this case, the transformation properties of the different curvature tensors proceeds easily
\begin{align}\label{scaletransR}
R_{\mu\nu} \rightarrow R_{\mu\nu}, \quad R \rightarrow e^{-2\sigma} R, \quad C^{\mu}_{~\nu\alpha\beta} \rightarrow C^{\mu}_{~\nu\alpha\beta} \ \ .
\end{align}
With these relations in hand, we can apply a scale transformation to the covariant action to recover the trace relation. We start with the quadratic action
\begin{align}\label{quadvar}
\delta_\sigma \overset{\scriptscriptstyle{(2)}}{\Gamma}[e^{2\sigma}g,A] = - \sigma\, \frac{b_i}{2} \int d^4x \sqrt{g} \, F_{\mu\nu} F^{\mu\nu} \ \ .
\end{align}
All terms with the form factor $1/\nabla^2$ are scale invariant, hence
\begin{align}
\delta_\sigma \overset{\scriptscriptstyle{(3)}}{\Gamma}_{pole}[e^{2\sigma}g,A]=0, \quad \delta_\sigma \overset{\scriptscriptstyle{(3)}}{\Gamma}_{ct.2}[e^{2\sigma}g,A] = 0
\end{align}
while terms with the form factor $\ln \nabla^2/\nabla^2$ cancel each other identically as described above
\begin{align}
\delta_\sigma \overset{\scriptscriptstyle{(3)}}{\Gamma}_{log}[e^{2\sigma}g,A] = - \delta_\sigma \overset{\scriptscriptstyle{(3)}}{\Gamma}_{ct.1}[e^{2\sigma}g,A] \ \ .
\end{align}
The anomalous trace of the energy-momentum tensor is easily determined from eq. (\ref{traceid}) and hence eq. (\ref{quadvar}) correctly reproduces the trace relation
\begin{align}
T_{\mu}^{~\mu} = \frac{b_i}{2} F_{\mu\nu} F^{\mu\nu} \ \ .
\end{align}
Again it is the logarithm which is the determining factor for the anomaly.

\section{Summary}\label{conc}

We have used a method which we refer to as {\em non-linear completion} in order to match the one-loop perturbative expansion of the QED effective action to a covariant expansion in the generalized curvatures. Within this procedure, the matching has been unique and relatively simple to implement. The results are given in eqs. (\ref{total}), (\ref{generalizedlog}) and (\ref{Weylaction}). These summarize the one-loop perturbative calculation involving one gravitational vertex.

The effective action also encodes the anomaly structure of the theory. For the anomaly, the important aspect is to generalize the feature that appeared as $\ln \Box$ in flat space. Our generalized result Eq. \ref{generalizedlog} contains many terms when expressed in terms of covariant derivatives and curvatures. All of these are required in order to both match the one loop perturbative calculation and to respect general covariance. There is also an interplay between these terms and various forms of scale and/or conformal invariance. There is a dispute in the literature about whether the anomaly comes from logarithmic terms or from the Riegert action\cite{Riegert}, e.g. see\cite{Schwimmer,Deser1,Erdmenger} and \cite{Giannotti,Mazur}. In our explicit computation, we showed that both forms are required in order for the action to respond properly to different types of transformation.

Given the simplicity of the perturbative result eq. (\ref{calc}), and the complexity of the expansion in the curvature eq. (\ref{generalizedlog}), one suspects that there is a better covariant representation for this result. However, the expansion in the curvature is one of the few covariant approximation schemes available and therefore needs to be well explored. We are not prepared to propose an improved representation in this paper, and are only trying to match the perturbative result to the standard form found when performing an expansion in the curvature. We (hopefully) reserve this improved representation to a future publication.

In addition, we note that some of the higher order terms in the curvature expansion have the potential to be singular in the infrared, and these higher order terms have only been lightly explored. More work is in progress to understand whether the non-local expansion in the curvature is useful in phenomenological applications \cite{Basem2}.

\section{\bf Acknowledgments}
We would like to thank A. Codello, S. Deser, E. Mottola, A. O. Barvinsky and G. Veneziano for useful discussions. This work has been supported in part by the U.S. National Science Foundation Grant No. PHY-1205896. Both authors acknowledge the hospitality and support of the Kavli Institute for Theoretical Physics during their stay for the Quantum Gravity Foundations: UV to IR workshop, during which some of this research was accomplished.

\appendix

\section{Momentum space representation}

In this appendix, we collect all the momentum space representations of the different curvature invariants. The quadratic density $\sqrt{g} F^2$ is very simple and we do not list here.
Moving to the cubic invariants, we find
\begin{align}
\int d^4x \sqrt{g} F_{\mu\nu} F^{\mu\nu} R = \int \frac{d^4p}{(2\pi)^4} \frac{d^4p^\prime}{(2\pi)^4} A^\alpha(p) A^\beta(-p^\prime) h^{\mu\nu}(-q) \, \mathcal{M}^{S}_{\alpha\beta\mu\nu} + \mathcal{O}(h^2)
\end{align}
where
\begin{align}
\mathcal{M}^{S}_{\alpha\beta\mu\nu} = 2 (p \cdot p^\prime \eta_{\alpha\beta} - p_\beta p^\prime_\alpha)(q_\mu q_\nu - q^2 \eta_{\mu\nu}) \ \ .
\end{align}
The invariant including the Ricci tensor reads
\begin{align}
\int d^4x \sqrt{g} F^\beta_{~\mu} F^{\alpha\mu} R_{\alpha\beta} = \int \frac{d^4p}{(2\pi)^4} \frac{d^4p^\prime}{(2\pi)^4} A^\alpha(p) A^\beta(-p^\prime) h^{\mu\nu}(-q) \, \mathcal{M}^{Ric}_{\alpha\beta\mu\nu} + \mathcal{O}(h^2)
\end{align}
where
\begin{align}
\nonumber
\mathcal{M}^{Ric}_{\alpha\beta\mu\nu} = \frac{1}{4} &p \cdot p^\prime \big[ (Q_\mu Q_\nu + q_\mu q_\nu) \eta_{\alpha\beta} - 2 (p^\prime_\mu p_\beta \eta_{\alpha\nu} + p^\prime_\nu p_\beta \eta_{\alpha\mu} + p^\prime_\alpha p_\mu \eta_{\beta\nu} + p^\prime_\alpha p_\nu \eta_{\beta\mu}) \\
& - q^2 (\eta_{\mu\alpha} \eta_{\nu\beta} + \eta_{\nu\alpha}\eta_{\mu\beta} + \eta_{\beta\alpha}\eta_{\mu\nu}) - 2 p_\beta p^\prime_\alpha \eta_{\mu\nu}\big] -\frac{1}{2} q_\mu q_\nu p_\beta p^\prime_\alpha \ \ .
\end{align}
Lastly, the invariant including the Riemann tensor reads
\begin{align}
\int d^4x \sqrt{g} F_{\alpha}^{~\beta} F^{\mu\nu} R^\alpha_{~\beta\mu\nu} = \int \frac{d^4p}{(2\pi)^4} \frac{d^4p^\prime}{(2\pi)^4} A^\alpha(p) A^\beta(-p^\prime) h^{\mu\nu}(-q) \, \mathcal{M}^{Riem}_{\alpha\beta\mu\nu} + \mathcal{O}(h^2)
\end{align}
where
\begin{align}
\nonumber
\mathcal{M}^{Riem}_{\alpha\beta\mu\nu} = \frac{1}{4} &\big[2 p^\prime_\alpha p_\beta (Q_\mu Q_\nu - q_\mu q_\nu) + q^4 (\eta_{\mu\alpha} \eta_{\nu\beta} +  \eta_{\mu \beta} \eta _{\nu\alpha}) \\
& + 2 q^2 (p_{\mu} p^{\prime}_{\alpha} \eta_{\nu\beta} + p_{\nu} p^{\prime}_{\alpha} \eta_{\mu\beta} + p_{\beta} p^{\prime}_{\nu} \eta_{\mu\alpha} + p_{\beta} p^{\prime}_{\mu} \eta_{\nu\alpha}) \big] \ \ .
\end{align}
In the above, $Q = p + p^\prime $ and $q= p - p^\prime$. Moreover, transversality and on-shellness are assumed
\begin{align}\label{transcond}
p^2 = p^{\prime 2} = 0, \quad p \cdot A(p) = p^\prime \cdot A(p^\prime) = 0 \ \ .
\end{align}

One can easily check that the matrix elements are both gauge-invariant and satisfiy energy-momentum conservation. For example,
\begin{align}
p^\alpha \mathcal{M}^{S}_{\alpha\beta\mu\nu} = p^{\prime \beta} \mathcal{M}^{S}_{\alpha\beta\mu\nu} = q^\mu \mathcal{M}^{S}_{\alpha\beta\mu\nu} =0\ \ .
\end{align}

Moreover, one can use the Weyl tensor given in eq. (\ref{weyl}) to get
\begin{align}
\int d^4x \sqrt{g} F_{\alpha}^{~\beta} F^{\mu\nu} C^\alpha_{~\beta\mu\nu} = \int \frac{d^4p}{(2\pi)^4} \frac{d^4p^\prime}{(2\pi)^4} A^\alpha(p) A^\beta(-p^\prime) h^{\mu\nu}(-q) \, \mathcal{M}^{C}_{\alpha\beta\mu\nu} + \mathcal{O}(h^2)
\end{align}
where
\begin{align}
\mathcal{M}^{C}_{\alpha\beta\mu\nu} = \frac{1}{6} (p \cdot p^\prime \eta_{\alpha\beta} - p_\beta p^\prime_\alpha) \left( q_\mu q_\nu - 3 Q_\mu Q_\nu - q^2 \eta_{\mu\nu}\right) \ \ .
\end{align}
The above tensor is clearly traceless as required since it stems from a conformally invariant Lagrangian.

\end{document}